\documentclass[12pt]{amsart}

\usepackage{geometry}                
\usepackage{natbib}
\usepackage{setspace}

\usepackage[utf8]{inputenc} 
\usepackage[T1]{fontenc}    
\usepackage{hyperref}       
\usepackage{url}            
\usepackage{booktabs}       
\usepackage{amsfonts}       
\usepackage{nicefrac}       
\usepackage{microtype}      
\usepackage{xcolor}         

\usepackage{amsmath}
\usepackage{amssymb}
\usepackage{amsthm}
\usepackage{todonotes}

\usepackage{tabularx}
\usepackage{threeparttable}

\usepackage[linesnumbered,ruled]{algorithm2e}
\SetKwInput{KwInput}{Input}                
\SetKwInput{KwOutput}{Output}              

\newtheorem{lemma}{Lemma}

\newtheorem{theorem}{Theorem}[section]

\newtheorem{assumption}{Assumption}

\renewcommand{\thelemma}{\arabic{lemma}}

\usepackage{cleveref}
\crefname{figure}{figure}{figures}
\creflabelformat{figure}{#2#1#3}

\crefname{equation}{equation}{equations}
\crefrangeformat{equation}{(#3#1#4) to~(#5#2#6)}
\creflabelformat{equation}{(#2#1#3)}

\crefname{lemma}{lemma}{lemmas}
\creflabelformat{lemma}{#2#1#3}
\crefname{design}{design}{designs}
\creflabelformat{design}{#2#1#3}
\crefname{proposition}{proposition}{propositions}
\creflabelformat{proposition}{#2#1#3}
\crefname{condition}{condition}{conditions}
\creflabelformat{condition}{#2#1#3}
\crefname{assumption}{assumption}{assumptions}
\creflabelformat{assumption}{#2#1#3}
\crefname{remark}{remark}{remarks}
\creflabelformat{remark}{#2#1#3}
\crefname{appendix}{appendix}{appendices}
\creflabelformat{appendix}{#2#1#3}

\renewcommand{\Pr}{\mathbb{P}}

\newcommand{\Exp}{\mathbb{E}}

\newcommand\raiseT[2]{\raisebox{0.25ex}{$#1#2$}}
\newcommand\tr{{\mathpalette\raiseT{\intercal}}}

\newcommand{\OR}{\mathrm{OR}}

\newcommand{\ORr}{\mathrm{OR_r}}

\newcommand{\IORp}{\widehat{\theta}_{p}}
\newcommand{\IORr}{\widehat{\theta}_{r}}

\newcommand{\AVARp}{\widehat{\sigma}^2_{p}}
\newcommand{\AVARr}{\widehat{\sigma}^2_{r}}

\newcommand{\SEp}{\widehat{\sigma}_{p}}
\newcommand{\SEr}{\widehat{\sigma}_{r}}

\newcommand{\Pml}{\widehat{\mathbb{P}}_{\mathrm{ML},k}}

\newcommand{\RR}{\mathrm{RR}}

\title[Average Adjusted Association]{Average Adjusted Association: Efficient Estimation with High Dimensional Confounders}

\author{Sung Jae Jun}
\address{Department of Economics,
Pennsylvania State University,
University Park, PA 16802}
\email{suj14@psu.edu}

\author{Sokbae Lee}
\address{Department of Economics,
Columbia University,
New York, NY, 10027}
\email{sl3841@columbia.edu}

\date{February 23, 2023}

\begin{document}

\maketitle

\begin{abstract}
The log odds ratio is a well-established metric for evaluating the association between binary outcome and exposure variables. Despite its widespread use, there has been limited discussion on how to summarize the log odds ratio as a function of confounders through averaging. To address this issue, we propose the Average Adjusted Association (AAA), which is a summary measure of association in a heterogeneous population, adjusted for observed confounders. 
To facilitate the use of it, we also develop efficient double/debiased machine learning (DML) estimators of the AAA. 
Our DML estimators use two equivalent forms of the efficient influence function, and are applicable in various sampling scenarios, including random sampling, outcome-based sampling, and exposure-based sampling. Through real data and simulations, we demonstrate the practicality and effectiveness of our proposed estimators in measuring the AAA.
\end{abstract}

\doublespacing

\section{INTRODUCTION}\label{sec:intro}

There are several statistical measures of association, among which the (log) odds ratio is one of the most popular ones. The odds ratio has been frequently used in medicine, biostatistics and epidemiology because it is simple, it has a natural interpretation in the standard logistic regression model, and it is invariant under various sampling designs that include case-control studies. See \citet{breslow1976regression,breslow1978there,breslow1980statistical} for early work and \citet{Bland1468,OR-JAMA} for how to use the odds ratio in practice. 

To describe the object of interest and the background more concretely, let $Y$ denote a binary outcome, $T$ a binary exposure and $X$ a vector of measured confounders/covariates. The conditional odds ratio between $Y$ and $T$ given $X=x$, which we denote by $\OR(x)$, provides a complete picture of the \emph{adjusted association} between $Y$ and $T$ for different subpopulations defined by different values of $x$. Conditioning on $X=x$ matters for two reasons: first, it is more plausible to make a causal interpretation out of $\OR(x)$ than the unconditional odds ratio between $Y$ and $T$ \citep[e.g.,][]{Holland:Rubin,greenland1999confounding}; second, the association can be heterogeneous over different subpopulations so that $\OR(x)$ is a complicated function of $x$ in general.

There are several approaches to modeling $\OR(\cdot)$. It is most common to 
parametrize the function $\OR(\cdot)$  typically via a logistic regression model, but it may suffer from misspecification. Therefore, effort has been made to mitigate the problem from the perspective of doubly robust estimation \citep[e.g.,][]{yun2007semiparametric,tchetgen2010doubly,tchetgen2013closed}. 
On the contrary, some authors have emphasized heterogeneity, advocating nonparametric estimation of the function $\OR(\cdot)$
\citep[e.g.,][]{chen2011nonparametric,hui2013nonparametric}.
However, fully nonparametric approaches generally suffer from the curse of dimensionality and they are often not a practical option in finite samples. In fact, all the numerical experiments in \citet{chen2011nonparametric} and \citet{hui2013nonparametric} are limited with the scalar $X$.

In this paper, we consider the case where the dimension of $X$ is high.  We introduce a new summary measure of association, which we call the \emph{average adjusted association}, by taking the average of conditional log odds ratios over covariates. That is, we think of  $\theta_0 := \Exp\{   \log\OR(X) \}$ as a summary measure of adjusted association {in a heterogeneous population}. The issues of what to condition on for $X$ and how to summarize $\OR(\cdot)$ have been recognized in the literature
\citep[e.g.,][]{muller2014estimating,greenland1999confounding}. However, to the best of our knowledge, {the average adjusted association has not been formally studied in the literature,} let alone how to deal with high dimensional $X$. {Indeed, many recently published papers in both natural and social sciences have used the odds ratio to report their findings but, to the best of our knowledge, they are all based on strong parametric assumptions that restrict heterogeneous association: a logistic regression model is the most popular, where heterogeneity in association needs to be pre-specified. For example, recent studies on determinants of callbacks to job applications \citep{farber2016determinants}, association of Alzheimer's dementia with genotypes \citep{reiman2020exceptionally}, and police use of force with respect to race \citep{hoekstra2022does} are all based on logistic regression and pre-specified odds ratios. It is worth noting that odds ratios are natural in these examples because callbacks, dementia, and use of force are all rare events that occur with small probabilities.}

{As we depart from parametric models, the odds ratio can be a complicated function of observed characteristics.} It is a natural and common practice to summarize a heterogeneous quantity by taking an average. For instance, when the treatment effect is heterogeneous, we frequently focus on the average treatment effect \citep[e.g.,][]{HIR:2003,ray2019debiased,Shi:Blei:Veitch} {as a summary parameter}. Despite the popularity and usefulness of the average treatment effect and related quantities, the literature on such a statistic based on the odds ratios is surprisingly sparse. {The only exception we are aware of is the Mantel-Haenszel approach that averages the (log) odds ratio across finite strata.} This paper fills this important gap.

Treating the function $\log\{\OR(\cdot)\}$ nonparametrically, we derive two equivalent forms of the efficient influence function for $\theta_0$, which we then explore to construct efficient double/debiased machine learning (DML) estimators of $\theta_0$. We take this approach because the class of DML estimators are generally more advocated than conventional plug-in efficient estimators particularly when $X$ is high-dimensional \citep[see, e.g.,][among many others]{chernozhukov2018double, lewis2021double}. The efficient influence function can be expressed in terms of either prospective probabilities (i.e., the probabilities of the outcome conditional on the exposure and confounders) or retrospective ones (i.e., the probabilities of the exposure given the outcome and confounders). Therefore, we have two types of DML estimators: they are asymptotically equivalent and efficient, but they are not the same in finite samples. Our work is the first to derive the forms of the efficient influence function and to propose suitable DML estimators of $\theta_0$. We also provide easy-to-follow computational and inferential algorithms for implementation.  {Given the basic nature of this research, we do not see potential negative societal impacts of our work, although we should mention that it generally requires much caution to draw causal inference from statistical association.}

\paragraph{Related Literature}

There are a couple of recent papers on automatic DML estimators (see, e.g. \citet{pmlr-v162-chernozhukov22a} and \citet{CNS:2022}). However, applying automatic orthogonalization to our setting is not necessarily better because (i) automation may induce additional approximation errors and (ii) the explicit formula obtained in this paper can provide useful insight into the estimation problem.

General theory for semiparametric efficiency is well-developed in the literature \citep[e.g.,][]{Ai:Chen:2012,ACHL:2014}. However, if we used the general framework, we would need to verify all the regularity conditions and our proof would not be self-contained. This type of verification is not needed for our setting because we can directly calculate the efficient influence function for $\theta_0$ with a more preliminary proof technique.


\paragraph{Replication Files}

The replication files for all the numerical results are available at \url{https://github.com/sokbae/replication-JunLee-2023-AISTATS}.

\section{AVERAGE ADJUSTED ASSOCIATION}\label{sec:parameters}

The odds ratio can be expressed by using either prospective or retrospective probabilities: i.e., for all $x$ in the support $\mathcal{X}$ of $X$,
\begin{align*}
&\OR(x)  \\
&:= 
	\frac{\Pr(Y=1 \mid T=1,X=x)}{\Pr(Y=1 \mid T=0,X=x)}\frac{\Pr(Y=0 \mid T=0,X=x)}{\Pr(Y=0 \mid T=1,X=x)}, \\
&=
	\frac{\Pr(T=1 \mid Y=1,X=x)}{\Pr(T=1 \mid Y=0, X=x)}\frac{\Pr(T=0 \mid Y=0,X=x)}{\Pr(T=0 \mid Y=1,X=x)},
\end{align*}
where the second equality is known as the invariance property of the odds ratio, which can be verified by the Bayes rule \citep[see, e.g.,][]{cornfield1951method}. The two expressions of $\OR$ can be used to develop two different machine learning estimators. 

Since the function $\OR$ is an infinite-dimensional object, it is generally difficult to estimate with high precision or even to communicate estimation results in a fully nonparametric manner, unless the dimension of $x$ is limited to 1 or 2.  In this context, we propose, as a scalar summary measure of association, to take the expectation of $\log\OR(\cdot)$ using the probability distribution of $X$. Specifically,  we define
\begin{equation}
	\theta_0 
    := 
    \Exp\{ \log\OR(X)  \},
\end{equation}
which can be understood as the \emph{Average Adjusted Association} (AAA) for the entire population. We have taken the logarithm before taking expectation because $\log\OR(x)$ corresponds to a coefficient in the traditional logistic model; for instance, if $\Pr(Y=1 \mid T=t,X=x) = G(\alpha_0 + \alpha_1 t + \alpha_2^\tr x)$, where $G(s) = \exp(s)/\{1+\exp(s) \}$, then $\log\OR(x) = \alpha_1$ for all $x\in \mathcal{X}$. But the essence of our results does not rely on this structure; all of our results can be straightforwardly modified to the case of aggregating without the logarithm. Since $\OR(\cdot)$ is a nonlinear function in general, $\theta_0$ differs from the log odds evaluated at the average value of $X$.

\section{EFFICIENT INFLUENCE FUNCTION}\label{sec:theory:eff}

We now characterize the efficient influence function for estimating $\theta_0$ and present two equivalent expressions. We first state {an assumption} that will be used throughout the paper. Recall that $Y$ and $T$ are binary variables, and let $\mathcal{X}$ be the support of $X$. 



\begin{assumption}\label{ass:bounded}
	There exists a constant $\epsilon>0$ such that for all $t,y\in \{0,1\}$ and for all $x\in \mathcal{X}$, we have $\epsilon \leq \Pr(Y=y, T=t \mid X=x) \leq 1-\epsilon$.
\end{assumption}

{\Cref{ass:bounded} is to ensure that all the four joint outcomes of $(Y,T)$ occur with positive probability conditional on any value of $X=x\in\mathcal{X}$. Therefore, conditioning on any outcome of $(Y,T)$ does not exclude any value of $X$ from the support: i.e., \cref{ass:bounded} implies that the joint support of $(Y,T,X)$ is given by $\{0,1\}\times\{0,1\}\times \mathcal{X}$. This requirement may not be trivial in some applications, unless we restrict out attention to a certain subpopulation. For example, if $Y$ represents prostate cancer, then it is reasonable to focus on the subpopulation of men. We are implicit about this type of (extra) conditioning throughout the analysis.}

{Also, under \cref{ass:bounded}, both of the prospective and retrospective representations of $\OR$ are well-defined for any $x\in \mathcal{X}$. Further, the existence of such an  $\epsilon>0$ in \Cref{ass:bounded} guarantees that $x \mapsto \OR(x)$ is uniformly bounded from below by zero and from above by infinity.}

{Below we discuss the efficient influence function for $\theta_0$ when \cref{ass:bounded} is the only restriction imposed on the distribution of $(Y,T,X)$.}

For $y, t \in \{0,1\}$, define
\begin{align*}
\Delta_{pt}(Y,T,X) 
&:=
\frac{ T^t (1-T)^{1-t}\{Y-\Pr(Y=1 \mid T=t,X)\}}{\Pr(Y=1 \mid T=t,X)\Pr(Y=0 \mid T=t,X)}, \\
\Delta_{ry}(Y,T,X) 
&:=
\frac{ Y^y (1-Y)^{1-y}\{T-\Pr(T=1\mid Y=y,X)\}}{\Pr(T=1 \mid Y=y,X)\Pr(T=0 \mid Y=y,X)}.
\end{align*}
Further, define
\begin{align}
F_p(Y,T,X) 
&:=
\log\OR(X) - \theta_0 + \frac{\Delta_{p1}(Y,T,X)}{\Pr(T=1\mid X)} - \frac{\Delta_{p0}(Y,T,X)}{\Pr(T=0\mid X)}, \label{pro-eff} \\
F_r(Y,T,X)
&:=
\log\OR(X) - \theta_0 + \frac{\Delta_{r1}(Y,T,X)}{\Pr(Y=1\mid X)} - \frac{\Delta_{r0}(Y,T,X)}{\Pr(Y=0\mid X)}. \label{ret-eff}
\end{align}
The efficient influence function for $\theta_0$ is given in the following theorem. 

\begin{theorem}\label{thm:main}
Suppose that \Cref{ass:bounded} holds. Then, the semiparametrically efficient influence function for $\theta_0$ is given by $F_p(Y,T,X) = F_r(Y,T,X)$. 
\end{theorem}

\Cref{thm:main} includes equality between $F_p(Y,T,X)$ and $F_r(Y,T,X)$; $F_p$ is based on the prospective expression of $\OR$, whereas $F_r$ uses the retrospective one. 
Theorem~\ref{thm:main} is proved in two steps: (i) by direct calculation, as in \citet{hahn1998role}, $\theta_0$ is pathwise differentiable along regular parametric submodels in the sense of \citet{newey1990semiparametric,newey1994asymptotic};
(ii) the pathwise derivative is an element of the tangent space, from which we can obtain the semiparametric efficiency bound $V_{\textrm{eff}}$ for $\theta_0$:
\begin{equation*}
V_{\textrm{eff}} := \Exp\{ F_p^2(Y,T,X)\} = \Exp\{ F_r^2(Y,Y,X)\}.
\end{equation*} 
The bound $V_{\textrm{eff}}$ can be achieved by double/debiased machine learning (DML) estimators based on the efficient influence function, i.e., the representation in either \eqref{pro-eff} or \eqref{ret-eff}. The DML approach has an advantage that it is robust to local perturbation on the unknown functions that need to be estimated in the first step, which is known as the Neyman orthogonality property \citep[see, e.g.,][]{chernozhukov2018double}.

In the remaining part of this section we formally show that the moment condition based on either \eqref{pro-eff} or \eqref{ret-eff} is indeed robust to local perturbation on the nonparametric components. For this purpose, note first that each of $F_p$ and $F_r$ depends on three nonparametric elements, i.e.,
\begin{align}
\eta_{p0}(x) &:= 
\begin{pmatrix}
\Exp(Y\mid T=0,X=x) \\
\Exp(Y\mid T=1,X=x) \\
\Exp(T\mid X=x) 
\end{pmatrix},
\label{eta-p0}  \\
\eta_{r0}(x) &:= 
\begin{pmatrix}
\Exp(T \mid Y=0,X=x) \\
\Exp(T\mid Y=1,X=x) \\ 
\Exp(Y\mid X=x) \bigr] 
\end{pmatrix},
\label{eta-r0} 
\end{align}
 respectively. Let $\mathcal{G}$ be a space of (measurable) functions on $\mathcal{X}$ such that $g\in \mathcal{G}$ satisfies $0<\inf g(x) \leq \sup g(x) < 1$: see \Cref{ass:bounded}.
 Let $\tilde F_p(\cdot)[Y,T,X]$ and $\tilde F_r(\cdot)[Y,T,X]$ denote the functionals defined on $\mathcal{G}^3$ such that $\tilde F_p(\eta_{p0})[Y,T,X] = F_p(Y,T,X)$ and $\tilde F_r(\eta_{r0})[Y,T,X] = F_r(Y,T,X)$. Then, Neyman orthogonality is concerned about the Gateaux derivatives of $\tilde F_p$ and $\tilde F_r$ at $\eta_{p0}$ and $\eta_{r0}$, respectively. 
\begin{theorem}\label{thm:neyman}
	Suppose that \Cref{ass:bounded} holds. Then,  
	the Gateaux derivative of $\tilde F_p(\cdot)[Y,T,X]$ at $\eta_{p0}$ has conditional mean zero given $X$ almost surely: i.e.,
	\begin{equation*}
	\Exp\Bigl[	\partial \tilde F_p \{\eta_{p0} +\gamma (\eta - \eta_{p0}) \} [Y,T,X]/\partial \gamma \Big|_{\gamma = 0}\  \Big| \ X \Bigr] 
	=
	0  \; \text{ a.s. }
	\end{equation*}
	for any direction $\eta$. The same is true for $\tilde F_r(\cdot)[Y,T,X]$ at $\eta_{r0}$ as well. 
\end{theorem}

\Cref{thm:neyman} says that both $F_p(Y,T,X)$ and $F_r(Y,T,X)$ provide Neyman orthogonal moments conditional on $X$.  One implication of the local robustness property is that the first step nonparametric estimation of $\eta_{p0}$ (or $\eta_{r0}$) in estimating $\theta_0$ will not have any first-order consequence, i.e., the limiting distribution would be the same as if $\eta_{p0}$ (respectively, $\eta_{r0}$) were known. In other words, all the adjustment terms that are needed to address the effect of the first step estimation are already reflected in $F_p$ (respectively, in $F_r$).

\section{DML ESTIMATORS}\label{sec:DML-est}

We now describe a couple of double/debiased machine learning (DML) estimators of $\theta_0$, for which we use the functions $F_p$ and $F_r$ as estimating equations.  We assume that a random sample $\{ (Y_i, T_i, X_i^\tr)^\tr:\ i=1,2,\dots, n\}$ is available, where $X_i$ is allowed to be high dimensional.

\subsection{Computational Algorithms}\label{sec:algorithms}

Let $K \geq 2$ be some fixed integer (say, 5, 10 or 20). For simplicity, assume that $n$ is divisible by $K$. Let $\{ I_k : k=1,\ldots,K \}$ denote a $K$-fold partition of $\{1,\ldots,n\}$ such that $| I_k | = n/K$ for each $k$. Suppose that one estimates 
all the conditional probabilities appearing in \eqref{pro-eff} or \eqref{ret-eff}, depending on which one to use for estimation, via machine-learning estimators  by using observations that belong to $I_k^c = \{1,\ldots,n\} \setminus I_k$ for each $k$. Using data in $I_k^c$ to estimate conditional probabilities evaluated at the points in $I_k$ is reminiscent of the traditional leave-one-out method.

In using the prospective formula in \eqref{pro-eff}, we start with
\begin{align*}
\hat p_{p,t,k}(x) 
&:=  
\Pml(Y=1 \mid T=t,X=x) \ \ \text{ for $t=0,1$}, \\
\widehat \OR_{p,k} (x)  
&:= 
\frac{\Pml(Y=1\mid T=1,X=x)}{\Pml(Y=0 \mid T=1,X=x)}\\
&\times \frac{\Pml(Y=0\mid T=0,X=x)}{\Pml(Y=1\mid T=0,X=x)}, \\
\widehat w_{p,k} (x) 
&:= \Pml(T=1 \mid X=x),
\end{align*}
where $\Pml$ denotes a machine-learning estimator of a probability model using observations that belong to $I_k^c$. We then define the prospective DML estimator $\IORp $ of $\theta_0$ by
\begin{align}\label{pDML-est}
 \IORp 
&:= 
\frac{1}{K} \sum_{k=1}^K  \frac{1}{| I_k |} \sum_{i \in I_k} \widehat\psi_{i,p,k} ,
\end{align}
where 
\begin{multline}\label{widehat_psi_pro}
\widehat\psi_{i,p,k} 
:=  
\log \widehat \OR_{p,k}(X_i)
+ \frac{T_i}{\widehat w_{p,k}(X_i)}  \frac{ \{ Y_i - \hat p_{p,1,k}(X_i) \}}{\hat p_{p,1,k}(X_i)\{ 1-\hat p_{p,1,k}(X_i) \}} \\
-
\frac{(1-T_i)}{\{1-\widehat w_{p,k}(X_i)\}}  
\frac{ \{Y_i - \hat p_{p,0,k}(X_i)\} }{\hat p_{p,0,k}(X_i)\{1-\hat p_{p,0,k}(X_i)\}}.
\end{multline}
The estimator $\hat \theta_p$ is asymptotically normal and efficient as we will show in \Cref{section:distribution}.  We have summarized the estimation procedure in \Cref{est-pHD}.

\begin{algorithm}[htbp]
 \KwInput{$\{ (Y_i, T_i, X_i): i=1,\ldots,n \}$, integer $K \geq 2$, machine learning methods for estimating probability models}
\KwOutput{estimate of $\theta_0$ and its standard error}

Construct a  $K$-fold partition $\{ I_k : k=1,\ldots,K \}$   of $\{1,\ldots,n\}$ of approximately equal size\;

For each $k$, use observations belonging to $I_k^c$ to obtain machine learning estimates  of 
$\Pr(Y=1 \mid T=1,X=x)$, $\Pr(Y=1 \mid T=0,X=x)$
and
$\Pr(T=1 \mid X=x)$, respectively\;

For each $k$, use observations belonging to $I_k$ to construct $\widehat\psi_{i,p,k}$ in \cref{widehat_psi_pro}\;

Obtain the estimate of $\theta_0$ by \cref{pDML-est}
and its standard error $\SEp /\sqrt{n}$ by 
\begin{align}\label{pDML-est-se}
 \AVARp  
&:= 
\frac{1}{K} \sum_{k=1}^K  \frac{1}{| I_k |} \sum_{i \in I_k} 
\Big\{ \widehat\psi_{i,p,k} - \IORp  \Big\}^2.
\end{align}

 \caption{Prospective DML estimator of $\theta_0$}\label{est-pHD}
\end{algorithm}

The retrospective DML estimator $\IORr $ of $\theta_0$ is defined analogously. That is, we start with
\begin{align*}
\hat p_{r,y,k}(x) 
&:=  
\Pml(T=1\mid Y=y,X=x) \ \ \text{ for $y=0,1$}, \\
\widehat \OR_{r,k} (x)  
&:= 
\frac{\Pml(T=1\mid Y=1,X=x)}{\Pml(T=0\mid Y=1,X=x)}\\
&\times \frac{\Pml(T=0\mid Y=0,X=x)}{\Pml(T=1\mid Y=0,X=x)}, \\
\widehat w_{r,k} (x) 
&:= \Pml(Y=1 \mid X=x),
\end{align*}
and we define
\begin{align}\label{rDML-est}
 \IORr
&:= 
\frac{1}{K} \sum_{k=1}^K  \frac{1}{| I_k |} \sum_{i \in I_k} \widehat\psi_{i,r,k},
\end{align}
where
\begin{multline}\label{widehat_psi_ret}
\widehat\psi_{i,r,k} 
:=  
\log \widehat \OR_{r,k}(X_i) 
+ \frac{Y_i}{\widehat w_{r,k}(X_i)}  \frac{ \{ T_i - \hat p_{r,1,k}(X_i) \}}{\hat p_{r,1,k}(X_i)\{ 1-\hat p_{r,1,k}(X_i) \}} \\
-
\frac{(1-Y_i)}{\{1-\widehat w_{r,k}(X_i)\}}  
\frac{ \{T_i - \hat p_{r,0,k}(X_i)\} }{\hat p_{r,0,k}(X_i)\{1-\hat p_{r,0,k}(X_i)\}}.
\end{multline}
The algorithm for the retrospective DML estimator can be stated easily by making simple modifications in \Cref{est-pHD}. We omit details for brevity.

Before we finish this subsection, we remark that the consistency of $\IORp$ and $\IORr$ does not require that $\widehat w_{p,k}$ and $\widehat w_{r,k}$ consistently estimate $w_p$ and $w_r$; indeed, $F_p(Y,T,X)$ and $F_r(Y,T,X)$ have mean zero even if $w_p$ and $w_r$ deviate from the truth. 

\subsection{Asymptotic Distributions}\label{section:distribution}

Let $\| \cdot \|_{P, 2}$ denote the $L_2(P)$-norm, where $P$ is the probability distribution of $(Y,T,X)$: i.e., $\| a \|_{P, 2} := \max_{1 \leq \ell \leq d} \left\{ \Exp [ a_\ell^2 (Y,T,X) ] \right\}^{1/2}$ for a $d$-dimensional vector-valued function $a := (a_1,\ldots,a_d)$. For each $k$, let 
\begin{equation*}
\widehat\eta_{n,p,k}(X) :=
\begin{pmatrix}
\Pml(Y=1\mid T=0,X) \\
\Pml(Y=1\mid T=1,X) \\
\Pml(T=1 \mid X)
\end{pmatrix}
\end{equation*}
denote the vector of machine learning estimators of $\eta_{p0}$ defined in \eqref{eta-p0},
using observations belonging to $I_k^c$. The first step estimators $\widehat\eta_{n,p,k}$ are inputs to the second step in \eqref{pDML-est}. Likewise, let $\widehat\eta_{n,r,k}$ be the vector of machine learning estimators of $\theta_{r0}$ defined in \eqref{eta-r0}, which will be used for the retrospective DML estimator of $\theta_0$.

\begin{assumption}[First-Stage Estimation]\label{ass:DML:est}
There exist sequences $\delta_n \geq n^{-1/2}$ and $\tau_n$ of positive constants both approaching zero such that for each $k=1,\ldots,K$, 
\begin{align*}
\| \widehat\eta_{n,p,k} - \eta_{p0} \|_{P, 2} &\leq \delta_n n^{-1/4}, \\
\| \widehat\eta_{n,r,k} - \eta_{r0} \|_{P, 2} &\leq \delta_n n^{-1/4},
\end{align*}
with probability no less than $1-\tau_n$.
\end{assumption}

\Cref{ass:DML:est} is a high-level assumption that may not be trivial if $X$ is high dimensional. For instance, it may fail even when all the conditional probabilities are logistically specified and they are estimated by the method of maximum likelihood if $X$ is high dimensional   \citep[e.g.,][]{sur2019modern,zhao2022asymptotic}. However, \Cref{ass:DML:est} is known to be attainable for a variety of machine learning methods. The primitive conditions for $\ell_1$-penalized logit estimators are worked out by \citet{vandegeer2008} and \citet{Belloni:2016:JBES} among others. A maximum likelihood approach with some adjustment for the dimensionality and signal strength of $X$ as in e.g., \citet{yadlowsky2021sloe} is another possibility although we do not pursue the latter in this paper.

An application of Theorems 3.1 and 3.2 of \citet{chernozhukov2018double} gives the following result that formally justifies the estimation and inference methods proposed in \Cref{sec:algorithms}.

\begin{theorem}\label{thm:DML}
Let $\{ \mathcal{P}_n: n \geq 1 \}$ be a sequence of sets of probability distributions of $(Y,T,X)$. 
Suppose that for all $n \geq 3$ and $P \in  \mathcal{P}_n$, 
\Cref{ass:bounded,ass:DML:est} hold and that 
	we have a random sample $\{(Y_i, T_i, X_i^\tr)^\tr: i=1,\ldots,n\}$. 
	Then, uniformly over $P \in  \mathcal{P}_n$, 
	\begin{equation*}
	\sqrt{n} \frac{ ( \IORp - \theta_0 ) }{\SEp} \rightarrow_d \mathbb{N} \left( 0, 1 \right) 
	 \text{ and } 
	\sqrt{n} \frac{ ( \IORr - \theta_0 ) }{\SEr} \rightarrow_d \mathbb{N} \left( 0, 1 \right).
	\end{equation*}
Furthermore,  both $\AVARp \rightarrow_p V_{\textrm{eff}}$ and $\AVARr \rightarrow_p V_{\textrm{eff}}$ uniformly over $P \in  \mathcal{P}_n$.
\end{theorem}

\Cref{thm:DML} establishes that both the prospective and retrospective DML estimators  are asymptotically normal and efficient. However, asymptotic equivalence does not imply that it is irrelevant which one to use between $\IORp$ and $\IORr$ in finite samples. In fact, the first steps of the two estimators involve different nonparametric regression functions, and therefore they generally lead to different estimates. Comparing the estimates and their standard errors can be a useful diagnostic check in practice; we recommend reporting inference results based on both estimators.

\section{FURTHER DISCUSSIONS}\label{sec:discussion}

\subsection{Nonparametric Estimation and Double Robustness}
Since we do not use any parametric specification to identify and estimate $\theta_0$, potential misspecification is not a concern, at least asymptotically{: it is a concern only to the extent that our choices of nonparametric estimators must satisfy \Cref{ass:DML:est}.} However, nonparametric approaches do rely on several input parameters, which are important for the performance of the estimators in finite samples. In this regard we show that there is a nonparametric version of double robustness.

For every $x\in \mathcal{X}$, we implicitly \emph{define} $\varphi_{p0}(x), \varphi_{r0}(x)$, and $\vartheta_0(x)$ by the following equations: for $t,y\in\{0,1\}$,
\begin{align}
\Pr(Y=1\mid T=t,X=x ) 
&= 
\frac{\exp\{ \varphi_{p0}(x) + t \vartheta_0(x)  \}}{1+\exp\{ \varphi_{p0}(x) + t \vartheta_0(x)   \}}, \label{eq:prop1}\\
\Pr(T=1\mid Y=y,X=x ) 
&= 
\frac{\exp\{ \varphi_{r0}(x) + y \vartheta_0(x)  \}}{1+\exp\{ \varphi_{r0}(x) + y \vartheta_0(x)  \}}. \label{eq:retro1}
\end{align}
For instance, \cref{eq:prop1} with $t=0$ defines $\varphi_{p0}(x)$. Here, we have four conditional probabilities to define three objects. This is because the four conditional probabilities are restricted by the invariance property of the odds ratio. Indeed, simple algebra shows that $\vartheta_0(x) = \log \OR(x)$, on which no restrictions have been imposed. 

Let $\mathcal{H}$ be the class of functions on $\mathcal{X}$, which contains $\varphi_{p0}, \varphi_{r0}$, and $\vartheta_0$. We then consider the function 
$m:\mathcal{H}^3\times \{0,1\}^2 \times \mathcal{X} \rightarrow \mathbb{R}$ defined by
\begin{align*}
m(\varphi_p, \varphi_r,\vartheta, Y, T,X) 
=
\{ Y - \Lambda_0(\varphi_p,X)  \}\{ T - \Lambda_0(\varphi_r,X)  \} \exp\{ - \vartheta(X) TY  \},
\end{align*}
where $\Lambda_0(\varphi, x) = \exp\{\varphi(x)\} / \bigl[ 1+\exp\{\varphi(x)\} \bigr]$. 

\begin{theorem}\label{thm:DR-ID}
	Suppose that \Cref{ass:bounded} holds. Then, for any $\varphi_p, \varphi_r \in \mathcal{H}$, we have
\begin{align*}
	\Exp\{ m(\varphi_{p0},\varphi_r,\vartheta_0,Y,T,X) \mid X\} 
	=
	\Exp\{ m(\varphi_p,\varphi_{r0},\vartheta_0,Y,T,X) \mid X\}
	=
	0 
	\quad a.s.
\end{align*}
	Further, for any $\vartheta \in \mathcal{H}$ and $x\in\mathcal{X}$ such that $\vartheta(x) \neq \vartheta_0(x)$,
	\[
	\Exp\{ m(\varphi_{p0},\varphi_{r0},\vartheta,Y,T,X) \mid X=x\}
	\neq 0 
	\quad a.s.
	\]
\end{theorem}

\Cref{thm:DR-ID} is a nonparametric extension of the double robustness idea of \citet{tchetgen2010doubly} and  \citet{tchetgen2013closed}: i.e., either  $\Pr(Y=1\mid T=0,X=x ) = \varphi_{p0}(x)$ or 
$\Pr(T=1\mid Y=0,X=x ) = \varphi_{r0}(x)$ (but not both) needs to be correctly specified to estimate  
$\vartheta_0(x) = \log\OR(x)$ consistently. Further, the second assertion shows that the function $m$ can be used to identify $\vartheta_0$ once $\varphi_{p0}$ and  $\varphi_{r0}$ are given by their definition. It can be shown that the efficiency bound for estimating $\theta_0 = \Exp\{ \vartheta_0(X) \}$ by using the conditional moment equation in \Cref{thm:DR-ID} is given by
\begin{align*}
 	F_{\textrm{DR}}(Y,T,X) 
 	=
 	\vartheta_0(X) - \theta_0 
 	+
 	\frac{1}{\Pr(Y=1, T=1 \mid X )}
 	\frac{  m(\varphi_{p0},\varphi_{r0},\vartheta_0,Y,T,X)}
 	{m(\varphi_{p0},\varphi_{r0},\vartheta_0,1,1,X) },
\end{align*}
which coincides with $F_p(Y,T,X) = F_r(Y,T,X)$. Therefore, we do not lose anything in terms of semiparametric efficiency in using the moment function $m$ to estimate $\theta_0$.

Therefore, it is possible to have an extra DML-based doubly robust algorithm for efficient estimation of $\theta_0$. However, we do not pursue this possibility in the current paper for a couple of reasons. As we described in \Cref{sec:DML-est}, the DML approach requires splitting the sample into multiple subsamples, but it often causes computational issues in practice when multiple tiers of nonparametric estimation are involved as in the current setup. Also, since our approach does not require any parametric specification at all, double robustness seems to have limited merits; there is no misspecification in the limit in the approach described in \Cref{sec:DML-est}.

\subsection{Interpretation: Association vs. Causation}\label{sec:interpretation}

One can rely on \Cref{thm:DML} to conduct statistical inference on $\theta_0$. For instance, using the prospective estimator $\hat \theta_p$, a (symmetric two-sided) $95\%$ confidence interval for $\theta_0$ can be obtained in the usual manner, i.e., $\hat\theta_p \pm 1.96\cdot \hat\sigma_p/\sqrt{n}$. Here, we emphasize that $\theta_0$ is understood as a simple association parameter to which we do not give any causal interpretation at this stage. However, $\theta_0$ can be used for causal inference under a few extra assumptions. 

In order to discuss causal inference, let $Y(t)$ denote the potential outcome when the treatment is exogenously fixed at $t$; i.e., the observed outcome $Y$ is equal to $Y(T) = Y(1)T + Y(0)(1-T)$. Then, $\OR(x)$ is related with the following causal parameters:
\begin{align*}
\vartheta_\OR(x)
&= 
\frac{\Pr\{ Y(1) = 1\mid X=x \}}{\Pr\{Y(1) = 0\mid X=x\}}\frac{\Pr\{ Y(0) = 0\mid X=x \}}{\Pr\{Y(0) = 1\mid X=x\}},\\
\vartheta_\RR(x) 
&= 
\frac{\Pr\{Y(1) = 1\mid X=x \}}{\Pr\{ Y(0) = 1\mid X=x \}}.
\end{align*}
That is, $\vartheta_\OR(x)$ represents the adjusted causal odds ratio, while $\vartheta_\RR(x)$ is the adjusted causal relative risk parameter. Below we summarize some facts that are known in the literature; see, e.g., \citet{Holland:Rubin} and \citet{jun2021causal} for more detail.

\begin{enumerate}
	\item (\textit{No confounding}) Suppose that there is no confounding conditional on $X=x$: i.e., for $t=0,1$, $Y(t)$ and $T$ are independent given $X=x$. Then, $\OR(x) = \vartheta_\OR(x)$. 
	\item (\textit{No confounding + MTR})
	If there is no confounding conditional on $X=x$ and the treatment is potentially beneficial but it never hurts, i.e., $Y(1)\geq Y(0)$ almost surely, which is termed as  the Monotone Treatment Response (MTR) assumption \citep[e.g.,][]{manski1997}, then $1\leq \theta_\RR(x) \leq \vartheta_\OR(x) = \OR(x)$, where the inequalities are sharp.
	\item (\textit{Confounding + MTR/MTS})
	 Suppose that there may be confounding even if we condition on $X=x$ but that the treatment is potentially beneficial (MTR). Further, suppose that those who have chosen to take the treatment have no smaller chance of ``success'' than those who have opted out, i.e., $\Pr\{ Y(t)=1\mid T=1,X=x\}\geq \Pr\{ Y(t) = 1\mid T=0, X=x\}$,  which is termed as  the Monotone Treatment Selection (MTS) assumption \citep[e.g.,][]{manski2000monotone}. Then, $1\leq \theta_\RR(x) \leq \OR(x)$, where the inequalities are sharp. 	
\end{enumerate}

Therefore, the average parameter $\theta_0$ and \Cref{thm:DML} can be used for causal inference on the entire population. For example, if the researcher is willing to assume that the treatment was randomly assigned conditional on $X$, then the usual symmetric confidence interval such as $\hat\theta_r\pm 1.96 \cdot\hat\sigma_r$ is a confidence interval for $\Exp\{ \vartheta_\OR(X)\}$. There is no general relationship between $\Exp\{ \vartheta_\OR(X)\}$ and $\Exp\{ \vartheta_\RR(X)\}$, but if $\Pr(Y=1)$ is close to zero 
(known as the rare-disease assumption), then the two causal parameters are known to be close to each other as long as there is no confounding given $X$. So, the symmetric confidence interval of $\theta_0$ can also be understood as an approximate confidence interval of $\Exp\{ \vartheta_\RR(X) \}$ in this scenario.

The no-confounding assumption or the rare-disease assumption can be unrealistic in some applications. For instance, many treatments of interest in social sciences such as education choices are deliberate decisions, and in such cases the no-confounding assumption is unrealistic. Even so, if one is willing to assume that education is potentially beneficial but it never hurts and that those who deliberately chose to take higher education is generally no less likely to ``succeed'' than those who did not, then $\theta_0$ can be understood as a sharp upper bound on $\Exp\{\vartheta_\RR(X)\}$. Therefore, a one-sided confidence interval such as $[1,\ \hat\theta_r + 1.64\cdot \hat \sigma_r]$ can be reported if the causal parameter $\Exp\{ \vartheta_\RR(X)\}$ is of interest.

\subsection{Averaging over a Subpopulation}

In practice there may be a subpopulation of particular interest, in which case averaging over the entire population may not provide the most relevant summary statistic. For example, consider the population of patients with a certain type of cancer. Suppose that $Y$ and $T$, respectively, indicate five-year survival (say, $Y=1$ for survival and $Y=0$ for death) and a certain type of treatment (say, $T=1$ for treatment and $T=0$ for no treatment). Here, it may be relevant to summarize the association between $Y$ and $T$ for those who received the treatment, i.e., $\theta_T(1) := \Exp\{\log\OR(X) \mid T=1  \}$. Alternatively, the association between $Y$ and $T$ for those who survived the cancer, which can be captured by $\theta_Y(1) := \Exp\{ \log\OR(X) \mid Y=1 \}$, can be an interesting quantity to look at. Of course, if the average adjusted association between $T$ and $Y$ is homogeneous across $X$, then there will be no difference among $\theta_0, \theta_T(1)$, and $\theta_Y(1)$. However, in general, they are all distinct and can be substantially different, depending on the degree of heterogeneity. 

Also, $\theta_T(1)$ and $\theta_Y(1)$ can be of interest if our access to a random sample is limited. For example, $\theta_T(1)$ is point identifiable even when we only have a treatment-based sample, where a half of the sample comes from the patients who received the treatment and the other half is from those who did not. Similarly, an outcome-based sample (e.g., case-control studies) is sufficient to identify $\theta_Y(1)$.  
The statistical analysis of $\theta_T(1)$ and $\theta_Y(1)$ is similar to that of $\theta_0$ and we do not repeat it here. 

In addition, there has been increasing interest in estimating individual level treatment effects \citep[e.g.,][]{pmlr-v70-shalit17a}. {Averaging over the entire population may have a risk of over-simplification: for example, if the association is positive for some values of $X$, and it is negative for other values of $X$, then the overall average association may be close to zero.} With this motivation in mind, suppose that there is a vector of low-dimensional confounders (say, $Z$) such that we are interested in estimating $z \mapsto \Exp\{\log\OR(X) \mid Z=z  \}$. Then it can be estimated by projecting $\log\OR(X)$ on a low-dimensional space of $Z$ as in  \citet{Ogburn:2015}, \citet{Lee:Okui:Whang}, and \citet{Semenov:Chernozhukov:20}.
However, it is a topic of future research to develop this idea formally.

\section{NUMERICAL STUDIES}\label{sec:application}


In this section, we provide numerical results to illustrate the usefulness of our approach. 


\subsection{Top Income and Higher Education}

We start with  a real-data example. 
Table \ref{tb:ACS-toy-example} summarizes data from 
American Community Survey (ACS) 2018, cross-tabulating 
the likelihood of top income by educational attainment. The sample is restricted to white males residing in California with at least a bachelor's degree. It is extracted from IPUMS USA  \citep{IPUMS}.
The ACS is an ongoing annual survey by the US Census Bureau that provides key information about the US population. 

\begin{table}[htp]
	\centering
	\caption{Top income and education \label{tb:ACS-toy-example}}
	\hspace{-3.5cm}
	\begin{tabularx}{11cm}{XcX}
	&	
	\begin{tabular}{crrr}
		\hline
		& \multicolumn{2}{c}{Beyond bachelor's} & Total \\
		Top income   & $T = 0$ &    $T = 1$ &  \\  \hline
		$Y = 0$ &    10,533 &   6,362 &  16,895   \\
		$Y = 1$ &    397 &   524 &    921   \\  
		Total        &  10,930 &  6,886 &  17,816   \\
		\hline
	\end{tabular}
	\end{tabularx}
\end{table}%

The binary outcome variable `Top income' ($Y$) is defined to be one if a respondent's annual total pre-tax wage and salary income is top-coded. In ACS 2018, the threshold income for top-coding is different across states. In our sample extract, the top-coded income bracket has median income \$565,000 and the next highest income that is not top-coded is \$327,000.  The binary exposure variable ($T$) is defined to be one if a respondent has a master's degree, a professional degree, or  a doctoral degree.

To adjust for individual differences, we 
include age and industry code as covariates ($X$).
In particular, cubic B-splines of age with 17 inner knots
as well as 254 industry dummies are included in this specification, which can be viewed as a high-dimensional setting.

Specifically, we implement $\ell_1$-penalized logistic estimation with \texttt{glmnet} package in R
\citep{glmnet}
 to estimate $\Pr(Y=1|T=t,X=x), t=0,1$ and $\Pr(T=1|X=x)$ for the prospective model
 (respectively, $\Pr(T=1|Y=y,X=x), y=0,1$ and $\Pr(Y=1|X=x)$ for the retrospective model)
 with 10-fold cross-fitting. The underlying assumption here is that the B-spline terms plus the industry dummies are rich enough to approximate $\Pr(Y=1|T=t,X=x)$ as well as $\Pr(T=1|X=x)$ for the prospective model
 (respectively, $\Pr(T=1|Y=y,X=x)$ as well as $\Pr(Y=1|X=x)$ for the retrospective model). The penalization tuning parameter is chosen by cross-validation  (that is, \texttt{lambda.min} in the \texttt{glmnet} package). 
Here, we focus on $\ell_1$-penalized estimators among other possible machine learning estimators because the primitive conditions for Assumption~\ref{ass:DML:est} are well established
for $\ell_1$-penalized logit estimators \citep[e.g.,][]{vandegeer2008,Belloni:2016:JBES}, as we mentioned in Section~\ref{section:distribution}.

\begin{table}[htbp]
	\centering
	\caption{Numerical results \label{tb:sieve-DML}}
		\begin{tabular}{lcc}
  \hline
Panel A: $\theta_0$ & Prospective & Retrospective \\ 
  \hline
Estimate & 0.72 & 0.71 \\ 
  Standard Error & (0.12) & (0.10) \\ 
   \hline
Panel B: $\exp(\theta_0)$ & Prospective & Retrospective \\ 
  \hline
Estimate & 2.06 & 2.04 \\ 
 95\% Confidence Interval & [1.61,2.63] & [1.67,2.49] \\ 
   \hline
\end{tabular}
\par
\end{table}%

\Cref{tb:sieve-DML} reports estimation results. 
Looking at Panel A, the prospective estimate of $\theta_0$ is 0.72, which is almost the same as the retrospective estimate of 0.71. In Panel B, we present point estimates of $\exp(\theta_0)$ and its confidence intervals  using asymptotic normality obtained in \Cref{thm:DML}. 

The estimates of $\exp(\theta_0)$ are comparable to the usual odds ratio in terms of its scale; 
therefore, they 
can be interpreted similarly. 
Furthermore, as can be seen from \Cref{tb:ACS-toy-example}, 
top income is a rare event (that is, the sample proportion of $\Pr(Y=1)$ is approximately $0.05$). When the outcome of interest is rare, an average of the conditional odds ratio approximates the average of conditional relative risk, namely the average of the ratio between $\Pr(Y=1|T=1,X)/\Pr(Y=1|T=0,X)$. Hence, obtaining a higher-level degree  is associated with doubling the chance of earning very high incomes. The 95\% confidence interval for the prospective estimate is $[1.61,2.63]$ (respectively, [1.67,2.49] for the retrospective model). As discussed in Section~\ref{sec:interpretation}, $\theta_0$ can be interpreted as the upper bound on the causal parameter $\Exp\{ \log \vartheta_\RR(X) \}$ if one assumes the MTR/MTS assumptions here.

Recall that the covariates consists of cubic B-splines of age  as well as industry dummies, resulting in 274 regressors. 
As $n = 17,816$, one may simply try to estimate a parametric logistic regression model with the same set of regressors. However, it turns out that this flexible parametric approach suffers from a couple of numerical issues:
(i)  a very small number of estimated coefficients are \texttt{NA} due to multicollinearity;
(ii) some of predicted probabilities are numerically zero. As a result, the conditional odds ratios are not defined for all values of the regressors.
To resolve these problems, we make some ad hoc adjustments: (i) we ignore the problematic regressors by setting their coefficients to be zero; (ii) we set a lower bound on the fitted probabilities. Specifically, any fitted probability less than 1e-6 is set to be 1e-6. Then, we estimate $\theta_0$  by simply plugging the fitted probabilities into the formula of $\theta_0 = \Exp\{ \log\OR(X)  \}$. The parametric plug-in estimates with the ad hoc adjustments turn out to be 0.38 (prospective estimate) and 0.88 (retrospective estimate). The large difference between the prospective and the retrospective estimates indicates that there is an anomaly in the plug-in estimates.
In addition, we also consider plug-in estimation of $\theta_0$ using $\ell_1$-penalized logistic estimation with the same specifications and tuning parameters as in DML estimation. Hence, in this case, the plug-in estimator is different from the DML estimator  in that (i) it uses a different estimating equation and (ii) it does not use cross-fitting. The resulting plug-in estimates are 0.78 (prospective estimate) and 0.68 (retrospective estimate). They look more similar to the DML estimators; however, there is no theoretically proven result regarding how to conduct inference with the $\ell_1$-penalized plug-in estimators. 


\subsection{A Monte Carlo Experiment}

We turn to a Monte Carlo experiment to make a more systematic comparison between the plug-in and DML estimators. We generate observations in the following way: (i) the covariates are randomly drawn from the empirical distribution of the ACS sample; 
(ii) the binary exposure variable is generated from a logit model with $\Pr(T=1|X) = G( \alpha_0 + 
\alpha_1 Age + \alpha_2 Age^2)$, where $G(\cdot)$ is the logit link function and the parameters $(\alpha_0, \alpha_1, \alpha_2)$ are chosen by fitting the logit model with the ACS sample;
(iii) the binary outcome variable is drawn from a logit model with 
$\Pr(Y=1|T,X) = G( \beta_0 + \beta_1 T + \beta_2 Age + \beta_3 Age^2)$, where 
the parameters $(\beta_0, \beta_1, \beta_2, \beta_3)$ are again chosen by fitting a logit model with the ACS sample. In this experimental design, the true model is such that $\theta_0 = \beta_1$ and the industry effects are null. However, we fit exactly the same specifications as in the previous real-data example to examine the differences between the plug-in and DML estimators. 
The only change made here is that 5-fold cross-validation is adopted for DML estimation to speed up Monte Carlo simulations. The sample size is $n=5,000$ and the number of Monte Carlo replications is 500.  

\begin{table}[htp]
	\centering
	\caption{Results of the Monte Carlo Experiment \label{tb:MC}}
	\hspace{-3.5cm}
	\begin{tabularx}{11cm}{XcX}
	&	
\begin{tabular}{lrrr}
  \hline
 Estimator & Mean  & Standard  & Size \\ 
  & Bias & Deviation & (10\%) \\  
  \hline
Prospective plug-in & 0.12 & 0.16 & NA \\ 
Retrospective plug-in  & 0.12 & 0.16 & NA \\ 
Prospective DML & 0.05 & 0.16 & 0.89 \\ 
 Retrospective DML & 0.09 & 0.16 & 0.84 \\ 
   \hline
\end{tabular}	\end{tabularx}
\end{table}%

Table~\ref{tb:MC} summarizes the results of the experiments. The prospective DML estimator has a much smaller mean bias than the prospective plug-in estimator without increasing the standard deviation. Further, its size (the probability of excluding the true value of $\theta_0$ in the confidence interval) is close to the 10\% nominal level. The retrospective DML estimator does not perform as well as the prospective DML estimator. This is due to the fact that experimental data are generated from a prospective logit model. Overall, the results of the experiment verify that the DML estimators are superior to the plug-in estimators when the underlying machine learning estimators are the $\ell_1$-penalized logistic regression estimators. 

\section{CONCLUSIONS}

Our proposed DML estimators offer a novel way of estimating the summary measure of association, namely the AAA functional $\theta_0$. In particular, we provide a method for statistical inference on $\theta_0$ based on asymptotic normality of our efficient DML estimators. 

This paper has focused on binary outcome and exposure. However, it is possible to define the AAA functional beyond the current setup. For example, following \citet{tchetgen2010doubly}, we can define the conditional odds ratio function as
\[
\mathrm{OR}(x):=\frac{f(y \mid t, x)}{f(y \mid t_0, x)} \frac{f\left(y_0 \mid t_0, x\right)}{f\left(y_0 \mid t, x\right)},
\]
where $Y$ and $T$ can take either discrete values, continuous values, or a mixture of both; $\left(y_0, t_0\right)$ is a user specified point in the sample space; and $f(y \mid t, x)$ is the conditional density of $Y$ given $T=t$ and $X=x$ with respect to a dominating measure $\mu$. It is a topic of future research to develop this idea formally.

\subsubsection*{Acknowledgements}

We would like to thank a meta-reviewer and four anonymous reviewers for helpful comments.

\appendix

      \renewcommand{\thelemma}{A.\arabic{lemma}} 
      \renewcommand{\theequation}{A.\arabic{equation}}
      \renewcommand{\thesection}{A}
      \setcounter{equation}{0}

\section*{Appendix}

\appendix


\section{Proofs of Theorems}

\noindent
\textbf{Proof of Theorem 3.1: }  The proof of  Theorem 3.1 has two parts.
We first establish the equivalence result in Theorem 3.1 and then show that $F_r(Y,T,X)$ is the efficient influence function of $\theta_0$.

\subsubsection*{Proof of the equivalence result in Theorem 3.1: }
The claim of $F_p(Y,T,X) = F_r(Y,T,X)$ can be verified by checking the four cases of $(Y,T)$ equal to $(0,0), (0,1), (1,0)$, or $(1,1)$. Below we will do this and show that
\begin{equation}\label{eq:equality}
\frac{\Delta_{p1}(Y,T,X)}{\Pr(T=1\mid X)} - \frac{\Delta_{p0}(Y,T,X)}{\Pr(T=0\mid X)}
=
\frac{\Delta_{r1}(Y,T,X)}{\Pr(Y=1\mid X)} - \frac{\Delta_{r0}(Y,T,X)}{\Pr(Y=0\mid X)}.
\end{equation}
In what follows we will use abbreviations like $P_{T|X}(t|x), P_{Y|TX}(y|t,x)$, etc., to denote $\Pr(T=t|X=x), \Pr(Y=y|T=t,X=x)$, and similar objects. Now, the left-hand side of \cref{eq:equality} can be expressed as
\begin{equation*}
TY a_{TY}(X) - T a_T(X) - Y a_Y(X) + a_o(X),
\end{equation*}
where
\begin{align*}
a_{TY}(x)
&=
\frac{1}{P_{T|X}(1|X)P_{Y|TX}(1|1,x)P_{Y|TX}(0|1,x)}
+
\frac{1}{P_{T|X}(0|x)P_{Y|TX}(1|0,x)P_{Y|TX}(0|0,x)},\\
a_T(x)
&=
\frac{1}{P_{T|X}(1|x)P_{Y|TX}(0|1,x)}
+
\frac{1}{P_{T|X}(0|x)P_{Y|TX}(0|0,x)},\\
a_Y(x)
&=
\frac{1}{P_{T|X}(0|x)P_{Y|TX}(1|0,x)P_{Y|TX}(0|0,x)},\\
a_o(x)
&=
\frac{1}{P_{T|X}(0|x)P_{Y|TX}(0|0,x)}.
\end{align*}
Similarly, the right-hand side of \cref{eq:equality} is
\begin{equation*}
TY b_{TY}(X) - T b_T(X) - Y b_Y(X) + b_o(X),
\end{equation*}
where
\begin{align*}
b_{TY}(x)
&=
\frac{1}{P_{Y|X}(1|X)P_{T|YX}(1|1,x)P_{T|YX}(0|1,x)}
+
\frac{1}{P_{Y|X}(0|x)P_{T|YX}(1|0,x)P_{T|YX}(0|0,x)},\\
b_Y(x)
&=
\frac{1}{P_{Y|X}(1|x)P_{T|YX}(0|1,x)}
+
\frac{1}{P_{Y|X}(0|x)P_{T|YX}(0|0,x)}, \\
b_T(x)
&=
\frac{1}{P_{Y|X}(0|x)P_{T|YX}(1|0,x)P_{T|YX}(0|0,x)},\\
b_o(x)
&=
\frac{1}{P_{T|X}(0|x)P_{T|YX}(0|0,x)}.
\end{align*}
Here, $a_o(x) = b_o(x)$ by the Bayes rule. Also, 
\begin{multline*}
a_Y(x)
=
\frac{P_{T|X}(0|x)}{P_{YT|X}(1,0|x)P_{YT|X}(0,0|x)}
=
\frac{P_{YT|X}(0,0|x)+P_{YT|X}(1,0|x)}{P_{YT|X}(1,0|x)P_{YT|X}(0,0|x)} \\
=
\frac{1}{P_{YT|X}(1,0|x)}
+
\frac{1}{P_{YT|X}(0,0|x)}
=
b_Y(X).
\end{multline*}
Similarly, $a_T(x) = b_T(x)$ and $a_{TY}(x) = b_{TY}(x)$ follows from simple algebra. \qed

\bigskip

Therefore, the proof of Theorem 3.1 will be complete if we show that $F_r(Y,T,X)$ is the efficient influence function of $\theta_0$. Before we do this, we prove several lemmas. Let $P_y(X) = \Pr(T=1\mid Y=y,X)$ and let $\tilde\gamma = (p, \gamma)^\tr$ be the parameter that denotes smooth regular parametric submodels, where $\gamma$ parametrizes the conditional likelihood
\[
\mathcal{L}_y(T,X) = f_{X|Y}(X\mid y) P_y(X)^T \{1-P_y(X)\}^{1-T}
\] 
and $p$ is the parameter whose true value is $p_0=\Pr(Y=1)$.  So, regular parametric submodels will be denoted by using $f_{X|Y}(x\mid y; \gamma)$ and $P_y(x; \gamma)$, along with the parameter $\gamma$. The truth will be denoted by $\tilde\gamma_0 = (p_0, \gamma_0)^\tr$. We will use the symbol $\partial_\gamma g(\gamma_0)$ to denote the derivative of the function $g$ with respect to $\gamma$ evaluated at $\gamma_0$. 

For $y=\{0,1\}$, define
\begin{equation*}
S_{X|Y}(X\mid y):= \partial_\gamma \log f_{X|Y}(X\mid y; \gamma_0) \quad \text{and} \quad
A_y(X) := \frac{\partial_\gamma P_y(X;\gamma_0)}{P_y(X)\{ 1- P_y(X) \}}.
\end{equation*} 
Note that $S_{X|Y}(X\mid y)$ is restricted only by $\Exp\{ S_{X|Y}(X\mid y) \mid Y=y \} = 0$, while the derivatives $\partial_\gamma P_y(X;\gamma_0)$ are unrestricted.

\begin{lemma}\label{lem:tangent}
	The tangent space is given by the set of functions of the form
	\begin{multline*}
	s(Y,T,X)
	=
	(1-Y)\bigl[ \tilde a_0(X) + \{ T- P_0(X)  \}\tilde b_0(X)   \bigr] 
	+
	Y\bigl[ \tilde a_1(X) + \{ T-P_1(X)  \} \tilde b_1(X)  \bigr] + \kappa(Y-p_0),
	\end{multline*}
	where $\kappa$ is a constant, and the functions $\tilde a_y$ and $\tilde b_y$ for $y=0,1$ are such that $\Exp\{ \tilde a_y(X)\mid Y=y\} = 0$ and $\Exp\{s^2(Y,T,X)\}<\infty$.
	
\proof
	The score along the regular parametric submodels at $\tilde\gamma_0$ can be expressed as $S(Y,T,X) = \bigl( S_p(Y,T,X)\ S_\gamma(Y,T,X) \bigr)^\tr$, where
	\begin{align}
	S_p(Y,T,X) 
	&= 
	\frac{Y-p_0}{p_0(1-p_0)}, \label{eq:score p}\\
	S_\gamma(Y,T,X) 
	&=
	Y S_{\gamma,1}(T,X) + (1-Y) S_{\gamma,0}(T,X), \label{eq:score gamma}
	\end{align}
	where $S_{\gamma,y}(T,X) = S_{X|Y}(X|y) + \{ T - P_y(X) \} A_y(X)$ for $y=0,1$. Noting that $S_{X|Y}(X\mid y)$ is only restricted by the conditional mean zero condition and $\partial_\gamma P_y(X;\gamma_0)$ is unrestricted, the lemma follows from linear combinations of $S_p(Y,T,X)$ and $S_\gamma(Y,T,X)$.  \qed		 
\end{lemma}

Consider $\theta_0(\tilde\gamma)$ defined by
\begin{equation*}
\theta_0(\tilde \gamma)
:=
\theta_Y(1;\gamma) p + \theta_Y(0;\gamma)(1-p),
\end{equation*}
where
\begin{equation*}
\theta_Y(y; \gamma)
=
\int_\mathcal{X} \log \ORr(x;\gamma) f_{X|Y}(x|y; \gamma) dx.
\end{equation*}

\begin{lemma}\label{lem:derivative}
	The derivatives of $\theta_0(\cdot)$ at $\tilde\gamma_0$ are given by
	\begin{equation*}
	\left\{
	\begin{aligned}
	\partial_p \theta_0(\tilde\gamma_0) &= \theta_Y(1) - \theta_Y(0),\\
	\partial_\gamma \theta_0(\tilde\gamma_0) &= \partial_\gamma \theta_Y(1;\gamma_0) p_0
	+
	\partial_\gamma \theta_Y(0;\gamma_0)(1- p_0),
	\end{aligned}
	\right.
	\end{equation*}
	where   
	\begin{equation} \label{eq:partial beta}
	\partial_\gamma\theta_Y(y;\gamma_0)
	=
	\int_\mathcal{X} \bigl\{ A_1(x) - A_0(x) + \log\ORr(x) S_{X|Y}(x\mid y) \bigr\} f_{X|Y}(x\mid y) dx
	\end{equation}

\proof
	It follows by direct calculation by using
	\begin{multline*}
	\partial_\gamma\ORr(x;\gamma_0)
	=
	\partial_\gamma P_1(x;\gamma_0) \frac{\{ 1-P_0(x)\}}{P_0(x)\{ 1-P_1(x)\}^2}
	- 
	\partial_\gamma P_0(x;\gamma_0) \frac{P_1(x)}{P_0^2(x)\{ 1-P_1(x)\}} \\
	=
	\bigl\{A_1(x) - A_0(x) \bigr\} \OR(x). \qed
	\end{multline*}
\end{lemma}

\subsubsection*{Proof of the influence function result in Theorem 3.1:}
We are now ready to complete the proof of Theorem 3.1; we will show that $F_r(Y,T,X)$ is the efficient influence function of $\theta_0$. First, note that
\begin{equation*}
	F_r(Y,T,X) 
	= 
	(Y-p_0)\{ \theta_Y(1) - \theta_Y(0)  \} 
	+
	Y F_{r1}(T,X)
	+
	(1-Y)F_{r0}(T,X),
\end{equation*}
where
\begin{align*}
F_{r1}(T,X) 
&:=
\log\ORr(X) - \theta_Y(1) +  \frac{1}{Q(X)} \frac{T-P_1(X)}{P_1(X)\{1-P_1(X) \}},\\
F_{r0}(T,X)
&:=
\log\ORr(X) - \theta_Y(0) - \frac{1}{1-Q(X)}\frac{T-P_0(X)}{P_0(X)\{1-P_0(X) \}},
\end{align*}
where $Q(X) := \Pr(Y=1|X)$.  Therefore, by \Cref{lem:tangent}, $F_r$ is clearly in the tangent space, and hence it suffices by \Cref{lem:derivative} to show that 
\begin{equation}\label{eq:goal1}
\Exp\{  F_r(Y,T,X) S_p(Y,T,X)  \} = \theta_Y(1) - \theta_Y(0), 
\end{equation}
and
\begin{multline} \label{eq:goal2}
\Exp\{  F_r(Y,T,X) S_\gamma(Y,T,X)  \} \\
=
\Exp\{ A_1(X) - A_0(X) \} 
+
p_0 \int_\mathcal{X} \log\ORr(x)S_{X|Y}(x\mid 1) f_{X|Y}(x\mid 1) dx \\
+
(1-p_0) \int_\mathcal{X} \log\ORr(x)S_{X|Y}(x\mid 0) f_{X|Y}(x\mid 0) dx,
\end{multline}
where $S_p$ and $S_\gamma = YS_{\gamma,1} +(1-Y)S_{\gamma,0}$ are provided in \cref{eq:score p,eq:score gamma}.  

First, by using the fact that $(Y - p_0) Y = (1-p_0) Y$ and $(Y-p_0) (1-Y)= -p_0(1-Y)$, we obtain 
\begin{align*}
F_r(Y,T,X) S_p(Y,T,X)
=
\frac{(Y-p_0)^2}{p_0(1-p_0)} \{ \theta_Y(1) - \theta_Y(0) \} 
+
\frac{Y}{p_0} F_{r1}(T,X)
-
\frac{1-Y}{1-p_0} F_{r0}(T,X),
\end{align*} 
of which the expectation shows that \cref{eq:goal1} is satisfied. 

Now, consider \cref{eq:goal2}. Note that 
\begin{multline}\label{eq:careful!}
F_r(Y,T,X) S_\gamma(Y,T,X)
=
S_\gamma(Y,T,X) (Y-p_0)\{ \theta_Y(1) - \theta_Y(0)\} \\
+
Y F_{r1}(T,X) S_{\gamma,1}(T,X)
+
(1-Y) F_{r0}(T,X) S_{\gamma,0}(T,X), 
\end{multline}
where the last two terms use the fact that $Y(1-Y) = 0$. Here, by using the fact that $(Y - p_0) Y = (1-p_0) Y$ and $(Y-p_0) (1-Y)= -p_0(1-Y)$ again, we obtain 
\begin{align} \label{eq:one}
\Exp\{ (Y-p_0) S_\gamma(Y,T,X)  \}
=
(1-p_0) \Exp\{ YS_{\gamma,1}(T,X)  \} 
-
p_0 \Exp\{ (1-Y)S_{\gamma,0}(T,X) \}
=
0,
\end{align}
where the last equality follows from $\Exp\{ S_{\gamma,y}(T,X) \mid Y=y \} = 0$. So, the first term on the right-hand side of \cref{eq:careful!} has been taken care of. 

For the second term on the right-hand side of \cref{eq:careful!}, note that
\begin{align}
&\Exp\{ Y F_{r1}(T,X) S_{\gamma,1}(T,X)  \}  \notag \\
&=
p_0 \Exp\Bigl[ \Bigl\{\log\ORr(X) + \frac{1}{Q(X)}\frac{T-P_1(X)}{P_1(X)\{1-P_1(X)\}}  \Bigr\} S_{\gamma,1}(T,X)  \  \Big| \ Y=1     \Bigr] \notag \\
&=
p_0 \Exp\Bigl[   
\log \ORr(X) S_{X|Y}(X\mid 1)
+
\frac{1}{Q(X)}\frac{\{T-P_1(X)\}^2}{P_1(X)\{1-P_1(X)  \}} A_1(X) \ \Big| \ Y=1
\Bigr] \notag \\
&=
p_0
\Exp\bigl\{   \log \ORr(X) S_{X|Y}(X\mid 1) \ \big| \ Y=1\bigr\}
+
\Exp\Bigl\{ \frac{Y A_1(X)}{Q(X)} \Bigr\} \notag \\
&=
p_0
\Exp\bigl\{   \log \ORr(X) S_{X|Y}(X\mid 1) \ \big| \ Y=1\bigr\} 
+
\Exp\bigl\{ A_1(X)\bigr\}, \label{eq:two}
\end{align}
where the last equality follows from the fact that $\Exp(Y|X) = Q(X)$.  Similarly, the expectation of the third term on the right-hand side of \eqref{eq:careful!} is 
\begin{align} \label{eq:three}
\Exp\{ (1-Y) F_{r0}(T,X) S_{\gamma,0}(T,X)  \} 
=
(1-p_0)\Exp\bigl\{   \log \ORr(X) S_{X|Y}(X\mid 0) \ \big| \ Y=0\bigr\} 
-
\Exp\bigl\{ A_0(X)  \bigr\}.
\end{align}
Combining \cref{eq:careful!} with \cref{eq:one,eq:two,eq:three} verifies \cref{eq:goal2}. So, we are done. \qed

\bigskip


\noindent
\textbf{Proof of Theorem 3.2: }  In the proof, we focus on 
the prospective estimating equation. 
The case of retrospective estimating equation is similar.
For simplicity, we suppress subscript $p$ in the notation. Recall that for $\eta = (a,b,c)^\tr \in \mathcal{G}^3$,
and
\begin{align*}
\tilde F( \eta )[Y,T,X]  
= 
\log \biggl[ \frac{b(X)\{ 1-a(X) \}}{\{1-b(X)\}a(X)} \biggr] - \theta_0 
+
\frac{T}{c(X)} 
\frac{\{ Y-b(X) \}}{b(X) \{ 1-b(X) \} }
-
\frac{1-T}{1-c(X)} 
\frac{\{ Y-a(X) \}}{a(X) \{  1-a(X)\} }
\end{align*}
and
$$
\eta_{0}(x) = \bigl( \Pr(Y=1|T=0,X=x), \Pr(Y=1|T=1,X=x), \Pr(T=1\mid X) \bigr)^\tr.
$$
Define
\begin{equation}\label{OR-def-proof}
 \OR (\eta) [X]
:=
 \frac{b(X)\{ 1-a(X) \}}{\{1-b(X)\}a(X)}.
\end{equation}	
Now, as in the proof of \cref{lem:derivative}, we have 
\begin{equation}\label{eq:dOR2}
\partial_\gamma \log \OR \big\{ \eta_0 + \gamma(\eta - \eta_0) \big\} [X] \big|_{\gamma = 0}
=
\frac{b(X) - b_0(X)}{b_0(x)\{1-b_0(x)\}}
-
\frac{a(X) - a_0(X)}{a_0(x)\{ 1-a_0(x)\}},
\end{equation}	
where $a_0(X) := \Pr(Y=1|T=0,X)$ and $b_0(X) := \Pr(Y=1|T=1,X)$.

Define
\begin{align*}
\Delta_0 (\eta) [Y,T,X]
&=
-
\frac{1-T}{1-c(X)} 
\frac{\{ Y-a(X) \}}{a(X) \{  1-a(X)\} },  \\
\Delta_1 (\eta) [Y,T,X]
&=
\frac{T}{c(X)} 
\frac{\{ Y-b(X) \}}{b(X) \{ 1-b(X) \} }. 
\end{align*}
Then, we have that 
\begin{align}
\Exp\Bigl( \partial_\gamma  \Delta_0\{\eta_0+\gamma(\eta-\eta_0)  \}[Y,T,X]\big|_{\gamma = 0} \ \Big|\ X \Bigr) 
&=
\frac{a(X) - a_0(X)}{a_0(X)\{  1-a_0(X) \} }, \label{Delta0eq} \\
\Exp\Bigl( \partial_\gamma \Delta_1 \{\eta_0+\gamma(\eta-\eta_0)  \}[Y,T,X]\big|_{\gamma = 0} \ \Big|\ X \Bigr)
&=
- \frac{b(X) - b_0(X)}{b_0(X)\{  1-b_0(X) \} }. \label{Delta1eq} 
\end{align}
Therefore, the conclusion follows from \cref{eq:dOR2,Delta0eq,Delta1eq}. \qed

\bigskip


\noindent
\textbf{Proof of Theorem 4.1: }
As in the previous proof, we focus on we focus on 
the prospective estimating equation. 
The case of retrospective estimating equation is similar.
As before, we suppress subscript $p$ in the notation.

We verify Assumptions 3.1 and 3.2 of Chernozhukov et al. (2018, C-DML hereafter). 
Following the notation used in C-DML, we have $\psi (W; \theta, \eta) = \tilde F_r( \eta )[Y,T,X]$ with $W = (Y, T, X)$:
so, our case belongs to that of linear scores, namely 
\begin{align*}
\psi (W; \theta, \eta) = 
\psi^a (W;  \eta) \theta
+  \psi^b (W;  \eta),
\end{align*}
where
\begin{align*}
\psi^a (W;  \eta) &= - 1, \\
\psi^b (W;  \eta) &=
\log \OR( \eta)[X] 
+
\frac{ T }{c(X)} 
\frac{  Y-b(X) }{b(X) \{ 1-b(X) \} }
-
\frac{ 1-T }{1-c(X)} \frac{  Y- a(X) }{a(X) \{  1-a(X)\} }
\end{align*}
and $\OR (\eta) [X]$ is defined in \eqref{OR-def-proof}.

\bigskip

\noindent
\emph{Verification of Assumption 3.1 of C-DML.}
Under Assumption 3.1, 
Specifically, 
Assumption 3.1 (a) of C-DML is satisfied by (3.2);
part (b) is by the linearity of the score $\psi$;
part (c) is by Assumption 3.1;
part (d) is by Theorem 3.2;
part (e) follows because $\Exp [  \psi^a (W;  \eta_0) ] = -1$.
 
\bigskip

\noindent
\emph{Verification of Assumption 3.2 (b) of C-DML.}
It holds trivially that $| \psi^a (W; \eta) |$ is bounded by a constant uniformly in $\eta$.
Moreover, by Assumption 3.1, there is a constant $c_1 < \infty$ such that 
\begin{align*}
| \psi (W; \theta, \eta) | \leq c_1
\end{align*}
uniformly in $\eta$ almost surely.

\bigskip

\noindent
\emph{Verification of Assumption 3.2 (d) of C-DML.}
Let $b_0(X) = \Pr(Y=1|X,T=1)$ and $c_0 = \Pr(T=1|X)$. Note that
\begin{align*}
\Exp [ \psi^2 (W; \theta, \eta_0)  ]
\geq 
\Exp
\Biggl[
\left\{ \log \OR(X) - \theta_0 \right\}^2
+
 \frac{1}{c_0(X)} \frac{  1 }{b_0(X) \{  1-b_0(X)\} } 
\Biggr],
\end{align*}
which is bounded from below by a constant under  Assumption 3.1.

\bigskip

Since Assumption 3.2 (a) of C-DML is the definition of the first stage estimator, 
Theorem 4.1 follows immediately from  Theorems 3.1 and 3.2 of C-DML, provided that we verify the remaining  Assumption 3.2 (c) of C-DML. 

\bigskip

\noindent
\emph{Verification of Assumption 3.2 (c) of C-DML.}
Using the notation used in C-DML, define
\begin{align*}
r_n &:= \sup_{\eta \in \mathcal{T}_N}  | \Exp [  \psi^a (W; \eta) - \psi^a (W; \eta_0) ] |, \\
r_n' &:= \sup_{\eta \in \mathcal{T}_N} ( \Exp [ | \psi (W; \theta, \eta) - \psi (W; \theta, \eta_0) |^2 ] )^{1/2}, \\
\lambda_n' &:= \sup_{\gamma \in (0,1), \eta \in \mathcal{T}_N}
|
\partial_\gamma^2 \Exp [ \psi (W; \theta, \eta_0 + \gamma(\eta - \eta_0)) ]  
|,
\end{align*}
where $\mathcal{T}_N \subseteq \mathcal{G}^3$ is a nuisance realization set that is discussed in detail in C-DML. 

\textit{Step 1.}
Note that $r_n = 0$ since  $\psi^a (W;  \eta)$ does not depend on $\eta$.

\noindent
\textit{Step 2.}
Now write that 
\begin{equation*}
\Bigl[ \Exp \bigl\{ | \psi (W; \theta, \eta) - \psi (W; \theta, \eta_0) |^2 \bigr\} \Bigr]^{1/2} 
= 
\| \psi (W; \theta, \eta) - \psi (W; \theta, \eta_0) \|_{P, 2} 
\leq
\| \mathcal{B}_1 \|_{P, 2}
+ 
\| \mathcal{B}_2 \|_{P, 2}
+ 
\| \mathcal{B}_3 \|_{P, 2}, 
\end{equation*}
where 
\begin{align*}
\mathcal{B}_1 
 &:=
\log \OR( \eta)[X] - \log \OR(X), \\
\mathcal{B}_2 &:=
\frac{T}{c(X)} 
\frac{ Y-b(X) }{b(X) \{ 1-b(X) \} }
-
\frac{T}{c_0(X)} 
\frac{ Y-b_0(X) }{b_0(X) \{ 1-b_0(X) \} }, \\
\mathcal{B}_3 &:=
\frac{ 1-T }{1-c(X)} \frac{  Y- a(X) }{a(X) \{  1-a(X)\} }
-
\frac{ 1-T }{1-c_0(X)} \frac{  Y- a_0(X) }{a(X) \{  1-a_0(X)\} }.
\end{align*}
Then, in view of Assumptions 3.1 and 4.1, there exists a sequence $\tilde{\delta}_n \rightarrow 0$ such that 
\begin{align*}
\Bigl[ \Exp \bigl\{ | \psi (W; \theta, \eta) - \psi (W; \theta, \eta_0) |^2 \bigr\}\Bigr]^{1/2}
&\leq
\tilde{\delta}_n
\end{align*}
holds with probability at least $1-\tau_n$.
This implies that we can take $r_n' =  \tilde{\delta}_n$.

\noindent
\textit{Step 3.}
Define 
$a_\gamma (X) := a_0(X) + \gamma \{ a(X)-a_0(X) \}$,
$b_\gamma (X) := b_0(X) + \gamma \{ b(X)-b_0(X) \}$
and
$c_\gamma (X) := c_0(X) + \gamma \{ c(X)-c_0(X) \}$.
Note that
\begin{equation*}
\partial_\gamma \log \OR\{\eta_0+\gamma(\eta-\eta_0)  \}[X]
=
\frac{ b(X)-b_0(X) }{b_\gamma (X) \{  1-b_\gamma (X)\} } 
- \frac{ a(X)-a_0(X) }{a_\gamma(X) \{  1-a_\gamma(X)\} }. 
\end{equation*}
In addition, 
\begin{align*}
& 
\partial_\gamma
\left[ \frac{ \{ Y- a_\gamma (X) \}}{a_\gamma (X) \{  1-a_\gamma (X)\} } \right] 
=
- \frac{ a(X)-a_0(X) }{a_\gamma(X) \{  1-a_\gamma(X)\} } 
- \frac{ \{  Y-a_\gamma(X)\} \{  1-2a_\gamma(X)\} }{a_\gamma^2(X) \{  1-a_\gamma(X)\}^2 } \{  a(X)-a_0(X)\}, \\
&
\partial_\gamma
\left[ 
\frac{ Y-b_\gamma(X) }{b_\gamma(X) \{ 1-b_\gamma(X) \} }
\right] 
=
- \frac{ b(X)-b_0(X) }{b_\gamma(X) \{  1-b_\gamma(X)\} } 
-
\frac{ \{  Y-b_\gamma(X)\} \{  1-2b_\gamma(X)\} }{b_\gamma^2(X) \{  1-b_\gamma(X)\}^2 } \{  b(X)-b_0(X)\}, \\
& 
\partial_\gamma
\left[ 
\frac{ T }{ c_\gamma(X) }
\right] 
=
-  
\frac{ T \{ c(X) - c_0(X) \} }{ \{ c_\gamma(X) \}^2 }, \\
&
\partial_\gamma
\left[ 
\frac{ 1-T }{ 1-c_\gamma(X) }
\right] 
= 
\frac{ (1-T) \{ c(X) - c_0(X) \} }{ \{ 1- c_\gamma(X) \}^2 }. 
\end{align*}

Combining these yields 
\begin{align*}
\partial_\gamma  \psi (W; \theta, & \eta_0 + \gamma(\eta - \eta_0))   
\\
&=
\frac{ b(X)-b_0(X) }{b_\gamma (X) \{  1-b_\gamma (X)\} } 
- \frac{ a(X)-a_0(X) }{a_\gamma(X) \{  1-a_\gamma(X)\} } 
\\
&
- \frac{ T }{ c_\gamma(X) }
\left[
\frac{ b(X)-b_0(X) }{b_\gamma(X) \{  1-b_\gamma(X)\} } 
+
\frac{ \{  Y-b_\gamma(X)\} \{  1-2b_\gamma(X)\} }{b_\gamma^2(X) \{  1-b_\gamma(X)\}^2 } \{  b(X)-b_0(X)\}
\right] 
\\
&
+ \frac{ 1-T }{ 1-c_\gamma(X) }
\left[
\frac{ a(X)-a_0(X) }{a_\gamma(X) \{  1-a_\gamma(X)\} } 
+ \frac{ \{  Y-a_\gamma(X)\} \{  1-2a_\gamma(X)\} }{a_\gamma^2(X) \{  1-a_\gamma(X)\}^2 } \{  a(X)-a_0(X)\}
\right] \\
&
- \frac{ T \{ c(X) - c_0(X) \} }{ \{ c_\gamma(X) \}^2 } 
\left[ 
\frac{ Y-b_\gamma(X) }{b_\gamma(X) \{ 1-b_\gamma(X) \} }
\right] 
- \frac{ (1-T) \{ c(X) - c_0(X) \} }{ \{ 1- c_\gamma(X) \}^2 }
\left[ \frac{ \{ Y- a_\gamma (X) \}}{a_\gamma (X) \{  1-a_\gamma (X)\} }
\right].
\end{align*}
If we take the second-order derivative in the equation above, we can see that each term of the second-order derivatives on the right-hand side can be bounded in absolute value by
 \begin{equation*}
 \chi(X)
 :=
 C_1 \{ a(X) - a_0(X) \}^2
 +
 C_2 \{ b(X) - b_0(X) \}^2
 +
 C_3 \{ c(X) - c_0(X) \}^2
 \end{equation*}
for some appropriate constants $C_1, C_2$, and $C_3$, because $\eta$ is in $\mathcal{G}^3$ and $|f(X) g(X)| \leq \{f^2(X) + g^2(X)\}/2$ for any real-valued functions $f$ and $g$. 
Therefore, by averaging $X$ out, we obtain 
\begin{align*}
\bigl|
\partial_\gamma^2 \Exp [ \psi (W; \theta, \eta_0 + \gamma(\eta - \eta_0)) ]  
\bigr|
\leq \Exp\{ \chi(X) \}.
\end{align*}
Then, by 
Assumption 4.1, there exists a sequence $\tilde{\delta}_n' \rightarrow 0$ such that 
\begin{align*}
\sup_{\gamma \in (0,1), \eta \in \mathcal{T}_N}
\bigl|
\partial_\gamma^2 \Exp [ \psi (W; \theta, \eta_0 + \gamma(\eta - \eta_0)) ]  
\bigr|
&\leq
\tilde{\delta}_n' n^{-1/2}
\end{align*}
holds with probability at least $1-\tau_n$.
Therefore, we can take $\lambda_n' =  \tilde{\delta}_n' n^{-1/2}$.
\qed

\bigskip


\noindent
\textbf{Proof of Theorem 5.1: }
We focus on the case, where $m$ is evaluated at $\varphi_{p0}, \vartheta_0$, and an arbitrary point $\varphi_r$; the other case is similar.
Recall that
\begin{align}
\Pr(Y=1\mid T=0,X) &= \frac{\exp\{\varphi_{p0}(X) \}}{1+\exp\{\varphi_{p0}(X) \}},\\
\Pr(Y=0\mid T=0,X) &= \frac{1}{1+\exp\{\varphi_{p0}(X) \}} 
\end{align}	
by the definition of $\varphi_{p0}$.  Therefore, 
\begin{align}
\Pr(Y=1\mid T,X)
&=
\frac{\exp\{\varphi_{p0}(X) \}\exp\{\vartheta_0(X)T\}}{1+\exp\{\varphi_{p0}(X) \}\exp\{\vartheta_0(X)T\}}  \label{eq:probY1}\\
&=
\frac{\Pr(Y=1\mid T=0,X) \exp\{\vartheta_0(X)T\}}{\Pr(Y=0\mid T=0,X)+\Pr(Y=1\mid T=0,X)\exp\{\vartheta_0(X)T\}}, \notag
\end{align}
where the second equality follows by dividing the numerator and the denominator by $1+\exp\{ \varphi_{p0}(X)\}$. Also,
\begin{align}
\Pr(Y=0\mid T,X)
&=
1-\Pr(Y=1\mid T,X) \label{eq:probY0}\\
&=
\frac{\Pr(Y=0\mid T=0,X)}{\Pr(Y=0\mid T=0,X)+\Pr(Y=1\mid T=0,X)\exp\{\vartheta_0(X)T\}}. \notag
\end{align}
So, we can combine \cref{eq:probY1,eq:probY0} by
\begin{equation}\label{eq:probY}
\Pr(Y=y\mid T,X)
=
\mathcal{D}^{-1}\Pr(Y=y\mid T=0,X) \exp\{\vartheta_0(X)Ty\},
\end{equation}
where
\begin{equation*}
\mathcal{D} := \Pr(Y=0 \mid T=0,X) + \Pr(Y=1 \mid T=0,X)\exp\{\vartheta_0(X)T\}.
\end{equation*}
Now, 
\begin{align*}
\Exp\{ m( \varphi_{p0},\varphi_r,\vartheta_0,Y,T,X ) \mid T,X\} 
=
\bigl\{ T - \Lambda_0(\varphi_r,X)  \bigr\}
\Exp\bigl[ \bigl\{ Y - \Lambda_0(\varphi_{p0},X)\bigr\} \exp\{-\vartheta_0(X) TY \}     \mid T,X \bigr],
\end{align*}
where the expectation factor on the right-hand side is equal to
\begin{multline*}
\sum_{y=0}^1 \{ y - \Lambda_0(\varphi_{p0},X)  \}\exp\{-\vartheta_0(X)Ty \} \Pr(Y=y\mid T,X) \\
=
\mathcal{D}^{-1}\sum_{y=0}^1\{ y - \Lambda_0(\varphi_{p0},X)  \} \Pr(Y=y\mid T=0,X) 
=
\mathcal{D}^{-1} \big[\Exp\{ Y\mid T=0,X \} - \Lambda_0(\varphi_{p0},X)\bigr]
=
0,
\end{multline*}
where the first equality is by \cref{eq:probY}. Therefore, the first assertion has been shown. For the second assertion, note that
\begin{align*}
\Exp\{ m(\varphi_{p0},\varphi_r,\vartheta, Y,T,X)   \mid T=0, X\} 
=
- \Lambda_0(\varphi_r,X) \Exp\bigl\{ Y-\Lambda_0(\varphi_{p0},X) \mid T=0,X \bigr\}
=
0.
\end{align*}
Further,
\begin{align*}
\Exp\{ m(\varphi_{p0},\varphi_r,\vartheta, Y,T,X)   \mid T=1, X\} 
=
\bigl\{ 1- \Lambda_0(\varphi_r,X)\bigr\} \Exp\bigl[ \{ Y-\Lambda_0(\varphi_{p0},X) \}\exp\{-\vartheta(X) Y  \} \mid T=1,X \bigr],
\end{align*}
where the expectation factor on the right-hand side is equal to
\begin{align*}
\sum_{y=0}^1 \{ &y - \Lambda_0(\varphi_{p0},X)\} \exp\{ -\vartheta(X)y \} \Pr(Y=y\mid T=1,X) \\
&=
\mathcal{D}^{-1}\sum_{y=0}^1 \{ y-\Lambda_0(\varphi_{p0},X)\} \exp\bigl[y \{\vartheta_0(X)-\vartheta(X)\} \bigr]  \Pr(Y=y\mid T=0,X) \\
&=
\frac{-\Lambda_0(\varphi_{p0},X) + \{1-\Lambda_0(\varphi_{p0},X) \}\exp\{\varphi_{p0}(X)\}\exp\{\vartheta_0(X)-\vartheta(X)\}   }{\mathcal{D}[1+\exp\{\varphi_{p0}(X)\}  ]} \\
&=
\frac{ \bigl[\exp\{\vartheta_0(X)-\vartheta(X)\} - 1 \bigr]\exp\{\varphi_{p0}(X) \} }{\mathcal{D}[1+\exp\{\varphi_{p0}(X)\}  ]^2},
\end{align*}
where the first equality is by \cref{eq:probY}. Then, we note that this is equal to zero if and only if $\vartheta(X) = \vartheta_0(X)$. \qed

\bibliographystyle{chicago}
\bibliography{AAA}

\end{document}